\documentclass[a4paper,oneside,final,notitlepage,onecolumn,12pt]{article}
\usepackage{amsfonts}
\usepackage{epsf}
\usepackage{graphicx}
\usepackage{amssymb,eso-pic}
\usepackage{latexsym}
\usepackage{tabularx}
\usepackage{amsxtra,accents}
\usepackage{hyperref}
\usepackage{t1enc}
\usepackage{amsmath}
\usepackage[T1]{fontenc}
\usepackage{amsmath}
\usepackage{authblk}

\usepackage{framed}
\usepackage{enumerate}
\usepackage{bm} 
\usepackage{mathrsfs}
\usepackage{amsthm}
\usepackage{latexsym}
\usepackage{tikz}
\usepackage{mathtools}
\DeclareGraphicsExtensions{.jpg,.jpeg,.pdf,.png,.eps}

\usepackage{ifpdf,epsfig,array,amsmath,amssymb}

\usepackage{psfrag}
\psfrag{na}{\footnotesize $n^a$}   
\psfrag{ta}{\footnotesize $\tau^a$}   
\psfrag{vna}{\footnotesize $\check{n}^a$}
\psfrag{vNa}{\footnotesize $\check{N}^a$}
\psfrag{scrW}{\footnotesize $\mathscr{W}_{\rho_0}$}
\psfrag{t=t1}{\footnotesize $\Sigma_{\tau_1}$}
\psfrag{t=t2}{\footnotesize $\Sigma_{\tau_2}$}
\psfrag{r=all}{\footnotesize $\rho=const$}
\psfrag{hna}{\footnotesize $\widehat{n}^a$}
\psfrag{tna}{\footnotesize $\tilde{n}^a$}
\psfrag{hab,Kab}{\footnotesize $h_{ab}, K_{ab}$}
\psfrag{thab,tKab}{\footnotesize $\tilde h_{ab}, \tilde K_{ab}$}
\psfrag{hhab,hKab}{\footnotesize $\widehat h_{ab}, \widehat K_{ab}$}
\psfrag{chab,cKab}{\footnotesize $\check{h}_{ab}, \check{K}_{ab}$}

\newcommand{\interior}[1]{\accentset{\smash{\raisebox{-0.12ex}{$\scriptstyle\circ$}}}{#1}\rule{0pt}{2.3ex}}
\definecolor{DGREEN}{rgb}{0,0.65,0.65}
\definecolor{grey1}{rgb}{0.52, 0.52, 0.51}

\usepackage[normalem]{ulem}
\usepackage{color}
\definecolor{blue}{rgb}{0,0,1}
\definecolor{red}{rgb}{1,0,0}

\DeclareFontFamily{OT1}{rsfs}{} \DeclareFontShape{OT1}{rsfs}{m}{n}{
<-7> rsfs5 <7-10> rsfs7 <10-> rsfs10}{}
\DeclareMathAlphabet{\mathscr}{OT1}{rsfs}{m}{n}

\def\sc{{\hskip 3.5pt {{}^{{}^{{}_{{}_{\bowtie}}}}} \kern -8.pt{}}}  
\def\SC{{\hskip 3.5pt {{}^{{}^{{}^{{}_{{}_{\bowtie}}}}}} \kern -10.5pt{}}}

\newtheorem{theorem}{Theorem}
\newtheorem{proposition}{Proposition}
\newtheorem{corollary}{Corollary}[section]
\newtheorem*{example*}{Example}
\newtheorem*{condition*}{Condition}
 
\usepackage[normalem]{ulem}
\usepackage{color}

\newcounter{mnotecount}

\newcommand{\mnotex}[1]
{\protect{\stepcounter{mnotecount}}$^{\mbox{\footnotesize $\bullet$\themnotecount}}$ 
	\marginpar{\color{red}
		\raggedright\tiny\em
		$\!\!\!\!\!\!\,\bullet$\themnotecount: #1} }

\begin{document}

\title{\vskip-1cm\textbf{On the construction of Riemannian three-spaces with smooth \\ inverse mean curvature foliation}\footnote{This is a written up version of a lecture given on 5th December 2019 at Institut Mittag-Leffler, Stockholm as part of the ongoing scientific program ``General Relativity, Geometry and Analysis: beyond the first 100 years after Einstein''. }}

\author[,1,2]{Istv\'an R\'acz \footnote{E-mail address:{\tt racz.istvan@wigner.hu}}}

\affil[1]{Faculty of Physics, University of Warsaw, Ludwika Pasteura 5, 02-093 Warsaw, Poland}

\affil[2]{Wigner RCP, H-1121 Budapest, Konkoly Thege Mikl\'{o}s \'{u}t  29-33, Hungary}

\maketitle

\begin{abstract}
Consider a one-parameter family of smooth Riemannian metrics on a two-sphere, $\mathscr{S}$. By choosing a one-parameter family of smooth lapse and shift, these Riemannian two-spheres can always be assembled into smooth Riemannian three-space, with metric $h_{ij}$ on a three-manifold $\Sigma$ foliated by a one-parameter family of two-spheres $\mathscr{S}_\rho$. It is shown first that we can always choose the shift such that the $\mathscr{S}_\rho$ surfaces form a smooth inverse mean curvature foliation of $\Sigma$. An integrodifferential expression, referring only to the area of the level sets and the lapse function, is also derived that can be used to quantify the Geroch mass.
If the constructed Riemannian three-space happens to be asymptotically flat and the $\rho$-integral of the integrodifferential expression is non-negative, then not only the positive mass theorem but, if one of the $\mathscr{S}_{\rho}$ level sets is a minimal surface, the Penrose inequality also holds. Notably, neither of the above results requires the scalar curvature of the constructed three-metric to be non-negative.
\end{abstract}
 


\section{Introduction}\label{introduction}
\setcounter{equation}{0} 

General relativity is a metric theory of gravity that makes it highly non-trivial to assign, in a sensible way, mass, energy, linear and angular momenta to bounded spatial regions. Yet, since the early seventies, it is also part of the common suppositions that the proper analytic characterization of highly energetic processes will be intractable unless suitable quasi-local, and possibly quasi-conserved, quantities can be found \cite{geroch, Penrose-1982, Christodoulou-Yau-1986}. 
The first important step towards the realization of these objectives was made by Geroch \cite{geroch} whose proposal greatly inspired most of the later developments. Geroch's quasi-local argument, in proving the positive mass theorem, assumed both the existence of an inverse mean curvature foliation and that the inspected timeslice is maximal \cite{geroch}. 
Since then, considerable progress had been made in proving the global existence of inverse mean curvature foliations \cite{HusIlm97, HusIlm01, Bray01, Bray09}. This is essential in getting a proof not merely of a quasi-local version of the positive mass theorem but also that of its close relative; the Penrose inequality \cite{geroch, JangWald77, Jang78, Kijowski-1986, Jezierski-Kijowski-1987, Jezierski-1994, Jezierski-1994b, Joerg01, Jezierski-Kijowski-2004, Szabados-2004, Mars-2009, Mars-2016}\,\footnote{For a more detailed account on the related references see,e.g.~\cite{Szabados-2004, Mars-2009, Mars-2016}.}. One of the most important steps forward was the proof of the Riemannian Penrose inequality by Huisken and Ilmanen \cite{HusIlm97, HusIlm01} (see also \cite{Bray01, Bray09}) which, under suitable conditions, also yields a quasi-local proof of the positive energy theorem.\footnote{It is important to be emphasized here that the positive energy theorem was proven originally in the late seventies by applying completely different techniques by Schoen, Yau and Witten \cite{Schoen-Yau-1979, Schoen-Yau-1979PRL, Schoen-Yau-1981, Witten-1981, Schoen-Yau-2017}.}  In proving the the Riemannian Penrose inequality, the scalar curvature of the Riemannian three-metric on the involved timeslices was assumed to be non-negative \cite{HusIlm97, HusIlm01}. Thereby, it is an interesting question on its own right if smooth inverse mean curvature foliations do exist whenever the non-negativity of the scalar curvature of the Riemannian three-metric is not guaranteed.  

\medskip

The main purpose of the present paper is to introduce a new method that enables us to construct a wide variety of Riemannian three-spaces. Each admits a smooth inverse mean curvature foliation and such that the scalar curvature is not required to be non-negative.  The original motivation for applying inverse mean curvature foliations was to prove that the Geroch mass is non-decreasing. The proposed new method provides an alternative construction of inverse mean curvature foliations. It is also generic as no field equations are used anywhere in the construction. To make this transparent, we start by revisiting generic variations of the Geroch mass. Note that in the conventional treatment of this variation, the shift of the flow, ``as unimportant'', was left out of the corresponding arguments. Nevertheless, as in our proposal, the shift vector acquires an important role, on good grounds, with the involvement of a non-trivial shift-vector field, which provides an additional motivation of recalling the argument of Geroch \cite{geroch}, in section \ref{sec: var-Geroch-energy}.

The basic ingredient of the proposed construction is a smooth one-parameter family of Riemannian two-metrics $\widehat{\gamma}_{AB}$ on a topological two-sphere $\mathscr{S}$, parameterized by $\rho$. In the simplest case, these two spheres can be assembled into a smooth three-manifold $\Sigma$ foliated by a one-parameter family of topological two-spheres. Accordingly, $\Sigma$ is expected (at least locally) to be a product space $\mathbb{R}\times\mathscr{S}$. (For more precise specifications, see section\,\ref{sec: prelim}.) At this kinematical level, in addition to the Riemannian two-metrics $\widehat{\gamma}_{AB}$ on the $\mathscr{S}_\rho$ level sets, we only have a flow that is lacing the $\mathscr{S}_\rho$ level sets into a three-manifold $\Sigma$.
The smooth Riemannian three-spaces, with metric $h_{ij}$, on this three-manifold are constructed by choosing suitable lapse and shift on $\Sigma$. 
The shift is chosen by solving an elliptic equation, \eqref{eq: shift_chi}, for one of its scalar potentials. This guarantees immediately---regardless of choice made for the other potential or the lapse---that the $\mathscr{S}_\rho$ level sets form a smooth inverse mean curvature foliation of $\Sigma$. Assuming that such a foliation has been fixed on $\Sigma$, a new integrodifferential expression, referring only to the area of the level sets and the lapse function, is also derived that can be used to quantify the Geroch mass. It is also shown that the quasi-local mass introduced by Bartnik \cite{Bartnik} in quasi-spherical foliations can be viewed as a special case of the Geroch mass. Besides, whenever the constructed Riemannian three-space happens to be asymptotically flat, and the $\rho$-integral of the aforementioned integrodifferential expression is non-negative, then the positive mass theorem holds. If, besides, one of the $\mathscr{S}_\rho$ level sets is a minimal surface in $\Sigma$, then the Penrose inequality is also guaranteed to be satisfied.

The proposed construction is distinguished from any of the former ones in that no {\it a priori} restriction is imposed on the scalar curvature of the constructed three-geometry. In particular, the constructed three-geometries will automatically accommodate non-maximal timeslices, which are out of the validity range of all the former discussions in \cite{HusIlm97, HusIlm01, Bray01, Bray09}.

\medskip

This paper is structured as follows: Section \ref{sec: prelim} is to recall the basic notions and notations. In section \ref{sec: var-Geroch-energy}, a careful inspection of generic variations of the Geroch mass---with the involvement of a non-trivial shift---is carried out. Subsection \ref{subsec: variation-Wrho} is to determine those conditions which guarantee monotonous behavior of the Geroch mass, whereas subsection \ref{subsec: getting-control} is to discuss the alternative ways of getting control on its $\rho$-dependence.  Section \ref{sec: IMCF} indicates that the proposed construction will indeed yield an inverse mean curvature foliation. The new method allowing the construction of a wide variety of Riemannian three-spaces is presented in detail in section \ref{sec: new-construction}. It starts by dynamical determination of the shift, in subsection \ref{subsec: dyn_shift}, and then proposes various choices for the lapse, in subsection \ref{subsec: dyn_lapse}. Asymptotically flat configurations are investigated in subsection \ref{subsec: more results}. The paper is closed in section \ref{sec: final-remarks} by our final remarks.

\section{Preliminaries}\label{sec: prelim}
\setcounter{equation}{0}

Consider a smooth, three-dimensional manifold $\Sigma$ endowed with a Riemannian metric $h_{ij}$.
Assume that $\Sigma$ is (almost everywhere) smoothly foliated by topological two-spheres. More precisely, we shall assume that there exists a (smooth) Morse function $\rho: \Sigma\rightarrow \mathbb{R}$ on $\Sigma$ that possesses only isolated non-degenerate critical points \cite{milnor, matsumoto}. The connected components of the $\rho=const$ level surfaces of this Morse function---which will also be signified by $\mathscr{S}_\rho$---are assumed to be topological two-spheres apart from level sets through the critical points. At a critical point of a Morse function $\rho: \Sigma\rightarrow \mathbb{R}$ the gradient $\partial_i\rho$ vanishes. A critical point is non-degenerate if the Hessian of $\rho: \Sigma\rightarrow \mathbb{R}$ is non-singular at that point, whereas the index of a critical point is the number of the negative eigenvalues of the Hessian. Note that the extreme index non-degenerate critical points, i.e.~with index zero or three, are distinguished as (locally) the $\mathscr{S}_\rho$ two-spheres shrink to a point while approaching those critical points, which signify (local) minimum or a maximum of $\rho: \Sigma\rightarrow \mathbb{R}$, respectively. Thereby, we shall refer to them as {\it origins}. Immediate trivial examples for three-manifolds fitting to the above requirements are the cylinders, disks, or three-spheres which are diffeomorphic to $\mathbb{R}\times \mathbb{S}^2$, $\mathbb{R}^3$ or $\mathbb{S}^3$, with no, one or two origins, respectively\,\footnote{The case of a cylinder,  $\mathbb{R}\times\mathbb{S}^2$, is self-explanatory.  `Disks', diffeomorphic to $\mathbb{R}^3$, can also be seen to be simple by referring to the Morse functions $\rho=\pm\sum_{i=1}^3 (x_i)^2$ with an index zero or three critical points, according to the choice made in the determination of $\rho$ for the ``$+$'' or ``$-$'' sign, at the origin of $\mathbb{R}^3$. The three-sphere, diffeomorphic to $\mathbb{S}^3$, is also simple with respect to the height function $\rho(x_1,x_2,x_3,x_4) \mapsto x_{4}$, where $\mathbb{S}^3=\{(x_1,x_2,x_3,x_4)\in\mathbb{R}^{4}\,|\, \sum_{i=1}^{4} (x_i)^2=\mathcal{R}^2 \}$ which possesses a pair of critical points with index three and zero at the `north and south poles', represented by the points $(0,0,0,\mathcal{R})$ and $(0,0,0,-\mathcal{R})$ in $\mathbb{R}^{4}$, respectively.}. 
It is worth emphasizing that generic (non-simple) three-manifolds, that are foliated (almost everywhere) by topological two-spheres, can always be given as the disjoint union of disks and cylinders that are glued together via $\rho=const$ slices through index one or two non-degenerate critical points of a Morse function $\rho: \Sigma\rightarrow \mathbb{R}$.\,\footnote{These are isolated pointlike pinches on the edges of `disks'. 
} For this reason, and to keep the arguments of the present paper simple enough, unless stated otherwise, we shall assume that $\Sigma$ is simple, i.e.
\begin{condition*} 
	$\Sigma$ is either a cylinder, a disk, or a three-sphere, respectively.
\end{condition*}

Even though a Morse function $\rho_M: \Sigma\rightarrow \mathbb{R}$ is chosen we may still be interested in relabeling the $\rho_M=const$ level surfaces by introducing a new radial coordinate $\rho=\rho(\rho_M)$. If the function $\rho=\rho(\rho_M)$ is strictly increasing, the orientation of the foliation, fixed by the Morse function, is preserved by the newly defined radial function. To allow the use of area-radial coordinates (for its definition, see subsection \ref{subsec: dyn_lapse}), the new radial function $\rho: \Sigma\rightarrow \mathbb{R}$ will only be assumed to be smooth apart from critical points. Hereafter, unless stated otherwise, the $\rho=const$ or $\mathscr{S}_\rho$ level surfaces will always refer to one of these generic types of radial coordinates, i.e.~we shall assume that a generic radial function $\rho: \Sigma\rightarrow \mathbb{R}$ has been fixed.
The transverse one-form field $\partial_i\rho$ is well-defined apart from critical points. With the help of the Riemannian metric $h_{ij}$ on $\Sigma$---apart from these isolated critical points---a \textit{unit form field} and a \textit{unit vector field} can be defined via the relations  $\widehat{n}_i=\partial_i\rho/(h^{kl}\partial_k\rho\,\partial_l\rho)^{1/2}$ and $\widehat{n}{}^i=h^{ij}\widehat{n}_j$, respectively. Both of these fields are \textit{normal} to the $\rho=const$ level surfaces. Using them the operator $\widehat{\gamma}{\,}^i{}_j=\delta{\,}^i{}_j-\widehat n{}^i\widehat n_j$, projecting fields to the tangent space of the level surfaces, gets also to be determined. 

\medskip

The intrinsic and extrinsic geometry of the $\mathscr{S}_\rho$ level surfaces can then be represented by the \textit{induced Riemannian two-metric} $\widehat{\gamma}{}_{ij}$ and the \textit{extrinsic curvature} $\widehat K_{ij}$, defined via the relations, 
\begin{equation}\label{hatextcurv}
\widehat{\gamma}{}_{ij}=\widehat{\gamma}{}^k{}_i\widehat{\gamma}{}^l{}_j h_{kl} \qquad {\rm and} \qquad \widehat K_{ij}= {{\widehat \gamma}^{l}}{}_{i} D_l\,\widehat n_j=\tfrac12\,\mathscr{L}_{\widehat n} {\widehat \gamma}_{ij}\,,
\end{equation}
respectively. Here $D_i$ is the covariant derivative operator associated with $h_{ij}$ and $\mathscr{L}_{\widehat n}$ denotes the Lie derivative with respect to the unit norm vector field ${\widehat n}{}^i$.  A $\rho=const$ level surface is referred to be \textit{mean-convex} if its mean curvature, ${\widehat  K}{}^l{}_{l}=\widehat{\gamma}{}^{ij}\widehat{ K}_{ij}=D_i {\widehat n}{}^i$, is positive everywhere on $\mathscr{S}_\rho$. 
 
\medskip 

Given a foliation $\mathscr{S}_\rho$ a vector field $\rho^i$ on $\Sigma$ is called to be a \textit{flow}, with respect to ${\mathscr{S}}_\rho$, if the integral curves of $\rho^i$---apart from the origins---intersect each level sets precisely once,  and also if $\rho^i$ is scaled such that $\rho^i\partial_i\rho=1$ where $\partial_i\rho$ is well-defined and non-vanishing. Note that such a flow is smooth, apart form the critical points, on $\Sigma$. It can always be chosen without referring to some background metric structure, and it is not unique as if a flow exists then infinitely many others do also exist. Whenever a Riemannian metric $h_{ij}$ is given, the flow $\rho^i$ can uniquely be characterized by its {\it lapse},  ${\widehat{N}}=(\widehat{n}{}^i\partial_i\rho)^{-1}$---which measures the normal separation of the surfaces ${\mathscr{S}_\rho}$---, and by its {\it shift},  $\widehat{N}{}^i=\widehat{ \gamma}{}^i{}_j\rho^j$, via the relation 
\begin{equation}\label{rhovf}
\rho^i=\widehat{N}\,\widehat{n}{}^i + {\widehat N}{}^i\,.
\end{equation}

\medskip

Notably, if the area of the foliating level sets is increasing a well-defined {\it quasi-local orientation of the ${\mathscr{S}_\rho}$ level sets} emerges. We may simply regard a flow $\rho^i$ {\it outward pointing} if the area of the level sets is increasing with respect to it  \cite{racz_bh_topology,racz_untrapped}. This happens, for instance, if the integral of the product $\widehat{N}{\widehat K}{}^l{}_{l}$ is greater than zero. To see this makes sense recall that the variation of the area $\mathscr{A}_\rho$ of the $\rho=const$ level surfaces, with respect to the flow $\rho^i$, reads as 
\begin{align}\label{eq: varA}
\mathscr{L}_{\rho} \mathscr{A}_\rho = {}&  \int_{\mathscr{S}_\rho} \mathscr{L}_{\rho}\, {\widehat{\boldsymbol{\epsilon}}}=  \int_{\mathscr{S}_\rho} \Big\{ \widehat{N}{\widehat K}{}^l{}_{l}+ \widehat{D}{}_i\widehat{N}{}^i \Big\}\, {\widehat{\boldsymbol{\epsilon}}}=  \int_{\mathscr{S}_\rho}  \widehat{N}{\widehat K}{}^l{}_{l} \,\, {\widehat{\boldsymbol{\epsilon}}}\,,
\end{align}
where, besides \eqref{rhovf}, the relations $\mathscr{L}_{{\widehat{n}}}\,{\widehat{\boldsymbol{\epsilon}}}={\widehat K}{}^l{}_{l}\,{\widehat{\boldsymbol{\epsilon}}}$ and $\mathscr{L}_{{\widehat{N}}}\,{\widehat{\boldsymbol{\epsilon}}}=\tfrac12\,{\widehat \gamma}^{ij}\mathscr{L}_{\widehat{N}} {\widehat \gamma}_{ij}\,{\widehat{\boldsymbol{\epsilon}}}
=\widehat{D}{}_i\widehat{N}{}^i\,{\widehat{\boldsymbol{\epsilon}}}$, along with the vanishing of the integral of the total divergence $\widehat{D}{}_i\widehat{N}{}^i$, were applied. 

Note that ${\widehat N}$  does not vanish on $\Sigma$ unless the Riemannian metric 
\begin{equation}
	h^{ij} 
	 = \widehat \gamma^{\,ij}+{\widehat N}^{-2}\big(\,\rho^{i}-\widehat  N{}^i\,\big)\big(\,\rho^{\,j}-\widehat{N}^j\,\big)
\end{equation}
gets to be singular. Since $\widehat{n}{}^i$ is a flow itself with ${\widehat N}\equiv 1$ and it is natural to require that the (quasi-local) orientations by $\widehat{n}{}^i$ and $\rho^i$ coincide, hereafter, we shall assume that ${\widehat N}$ is positive throughout $\Sigma$. Combining the foregoing we get that if the integral of ${\widehat{N}}{\widehat K}{}^l{}_{l}$ is greater than zero, the area is, indeed, increasing with respect to $\rho^i$ and, in turn, that the flow $\rho^i$ may be referred outward pointing. The above argument can also be used to verify that for mean-convex surfaces the area is ``piece-wise strictly increasing'' as not only the total area but the area of any local surface element is increasing with respect to {\it outward pointing} flows. 

\section{Variation of the Geroch mass}\label{sec: var-Geroch-energy}

In proceeding the variation of the (quasi-local) Geroch mass\,\footnote{In most of the cases \eqref{eq: GQLM} is referred to as the (Riemannian)  Hawking mass in spite of the fact that the Geroch and  Hawking quasi-local mass differ conceptually (see, e.g.\,\cite{Szabados-2004}). For instance, the Hawking mass is known to depend, beside on the geometry of two-surface $\mathscr{S}_\rho$ within $\Sigma$, also on the way $\Sigma$ is embedded into an ambient space. As opposed to this the Geroch mass depends only on the geometry of $\mathscr{S}_\rho$ within $\Sigma$. In particular the Geroch mass is always smaller than or equal to the Hawking mass, and they are known to be equal only if for the extrinsic curvature $K_{ij}$ of $\Sigma$, defined with respect to an ambient space, the contraction $\widehat{\gamma}^{ij}K_{ij}$ vanishes on $\mathscr{S}_\rho$ \cite{Szabados-2004}. Based on these observations it is preferable to distinguish these two concepts and hereafter we shall refer to \eqref{eq: GQLM} as the Geroch mass.} \cite{geroch}
\begin{equation}\label{eq: GQLM}
	{M}_{\mathcal{G}} = \frac{\mathscr{A}_\rho^{1/2}}{64 \pi^{3/2}}\int_{\mathscr{S}_\rho} \left[\, 2\,{\widehat R} - ({\widehat K}{}^l{}_{l})^2 \,\right]{\widehat{\boldsymbol{\epsilon}}} \,,
\end{equation}
will play central role.
It is straightforward to see that the generic variation of the Geroch mass can be given as 
\begin{equation}\label{eq: gen-var-GE}
\mathscr{L}_{\rho} {M}_{\mathcal{G}} =  \frac{\mathscr{A}_\rho^{1/2}}{64 \pi^{3/2}}\, \Big[\,\mathscr{L}_{\rho} W + \tfrac12\,\mathscr{L}_{\rho} (\log \mathscr{A}_\rho)\cdot W \,\Big]\,,
\end{equation}
where $W=W(\rho)$ stands for the pure integral term in \eqref{eq: GQLM}, 
\begin{equation}\label{Wrho}
W = \int_{\mathscr{S}_\rho} \left[\, 2\,{\widehat R} - ({\widehat K}{}^l{}_{l})^2 \,\right]{\widehat{\boldsymbol{\epsilon}}}\,.
\end{equation}

If the non-decreasing of the Geroch mass was guaranteed and for some $\rho=\rho^*$ value the integral ${M}_{\mathcal{G}}(\rho^*)$ was zero or positive then ${M}_{\mathcal{G}}\geq 0$ would automatically hold to the exterior of $\mathscr{S}_{\rho^*}$ in $\Sigma$. 

As an important special case, it is worth mentioning that ${M}_{\mathcal{G}}$ vanishes at {\it regular origins}. A point $p\in \Sigma$ was considered to be an {\it origin} if it was an isolated non-degenerate index zero or three critical point of the Morse function $\rho: \Sigma \rightarrow \mathbb{R}$. Replacing the Morse radial function by the area-radial coordinate---for the precise definition of the latter see subsection \ref{subsec: dyn_lapse}---and using coordinates, $(\rho,x^A)$, adapted to the foliation and the flow, in a neighborhood of $p$ the notion of regular origin can be introduced as follows. An origin at $p$ is considered to be {\it regular} if there exist smooth bounded fields $\widehat N{}_{(2)}, \widehat N^A_{(1)}$  and $\widehat{\gamma}{}_{AB}^{\,(4)}$ such that in a neighborhood of $p$ on $\Sigma$ the relations 
\begin{align}\label{eq: origin} 
\hskip-0.3cm\hat N=1+(\rho-\rho^*)^2\,\hat N{}_{(2)}\,, \  \widehat N^A= (\rho-\rho^*)\,\widehat N^A_{(1)}\,, \ \widehat{\gamma}{}_{AB} = (\rho-\rho^*)^2\,{\interior \gamma}_{AB} + (\rho-\rho^*)^4\,\widehat{\gamma}{}_{AB}^{\,(4)}
\end{align}
hold, where ${\interior \gamma}_{AB}$ stands for the unit two-sphere metric. On the one hand, these conditions exclude the occurrence of a conical singularity at $p$, whereas, on the other hand, they also guarantee that the integrand in \eqref{Wrho} tends to zero while $\rho \rightarrow \rho^*$. This, along with the fact that near $p$, up to leading order, $\mathscr{A}_\rho$ is proportional to $(\rho-\rho^*)^2$, implies then that the Geroch mass vanishes in the $\rho \rightarrow \rho^*$ limit, i.e.~at regular origins.  
Hereafter we shall assume that if an origin occurs on $\Sigma$, it is also regular.

\subsection{The variation of $W(\rho)$}\label{subsec: variation-Wrho}

We proceed by deriving the generic variations of $W(\rho)$. In doing so the key equation we shall apply relates the scalar curvatures of $h_{ij}$ and ${\widehat \gamma}_{ij}$ via
\begin{equation}\label{generic}
{}^{{}^{(3)}}\hskip-1mm R= {\widehat R} - \left\{2\,\mathscr{L}_{\widehat n} ({{\widehat K}^l}{}_{l}) + ({{\widehat K}^{l}}{}_{l})^2 + \widehat K_{kl} {\widehat K}^{kl} + 2\,{\widehat N}^{-1}\,{\widehat D}{}^l {\widehat D}_l {\widehat N} \right\}\,.
\end{equation} 
Note that this equation can be deduced from the Gauss-Codazzi relations (see, e.g. (A.1) in \cite{racz_geom_det}); thereby, it holds on $\Sigma$ without referring to any field equations.

\medskip

By varying $W(\rho)$ with respect to an arbitrary flow we get 
\begin{align}\label{propW}
\mathscr{L}_{\rho} W = {}&  - \int_{\mathscr{S}_\rho} \mathscr{L}_{\rho}\Big[\,\big({\widehat K}{}^l{}_{l}\big)^2\, {\widehat{\boldsymbol{\epsilon}}}\,\Big] = - \int_{\mathscr{S}_\rho} \Big\{ \widehat{N} \, \mathscr{L}_{\widehat{n}}\Big[\,\big({\widehat K}{}^l{}_{l}\big)^2\, {\widehat{\boldsymbol{\epsilon}}}\,\Big]+\mathscr{L}_{\widehat{N}}\Big[\,\big({\widehat K}{}^l{}_{l}\big)^2\, {\widehat{\boldsymbol{\epsilon}}}\,\Big]\Big\}
\nonumber \\
= {}&  - \int_{\mathscr{S}_\rho} \big({\widehat{N}} {\widehat K}{}^l{}_{l}\big) \left[\, 2\,\mathscr{L}_{{\widehat{n}}}\,({\widehat K}{}^l{}_{l}) +  ({\widehat K}{}^l{}_{l})^2 \right]{\widehat{\boldsymbol{\epsilon}}} - \int_{\mathscr{S}_\rho}\widehat{D}{}_i\big[\big({\widehat K}{}^l{}_{l}\big)^2\widehat{N}{}^i\,\big]    {\widehat{\boldsymbol{\epsilon}}} \nonumber \\
= {}& - \int_{\mathscr{S}_\rho} \big({\widehat{N}} {\widehat K}{}^l{}_{l}\big) \left[\, (\, {\widehat R} - \hskip-1mm{}^{{}^{(3)}}\hskip-1mm R \,) - \widehat K_{kl} {\widehat K}^{kl} -  2\,{\widehat N}^{-1}\,{\widehat D}{}^l {\widehat D}_l {\widehat N} \,\right]{\widehat{\boldsymbol{\epsilon}}} \,,
\end{align}
where on the first line \eqref{rhovf} and the Gauss-Bonnet theorem, on the second line again the relations $\mathscr{L}_{{\widehat{n}}}\,{\widehat{\boldsymbol{\epsilon}}}=({\widehat K}{}^l{}_{l})\,{\widehat{\boldsymbol{\epsilon}}}$ and $\mathscr{L}_{{\widehat{N}}}\,{\widehat{\boldsymbol{\epsilon}}}
=(\widehat{D}{}_i\widehat{N}{}^i)\,{\widehat{\boldsymbol{\epsilon}}}$, whereas on the third line \eqref{generic} and the vanishing of the integral of $
\widehat{D}{}_i\big[\big({\widehat K}{}^l{}_{l}\big)^2\widehat{N}{}^i\,\big] $
were used.  
Applying then the Leibniz rule we get that 
\begin{equation}\label{eq: DivNhat}
{\widehat N}^{-1}{\widehat D}{}^l {\widehat D}_l {\widehat N}= {\widehat D}{}^l \big({\widehat N}^{-1}{\widehat D}_l {\widehat N} \big) + {\widehat N}^{-2}\,{\widehat \gamma}^{kl}\,({\widehat D}{}_k {\widehat N}) ({\widehat D}_l {\widehat N})\,,
\end{equation}
and---by introducing the trace-free part of ${\widehat K}_{ij}$ as $\interior{{\widehat K}}_{ij} = {\widehat K}_{ij} -  \tfrac12\,{\widehat \gamma}_{ij} \,({\widehat K}{}^l{}_{l})$---we also get that
\begin{equation}\label{eq: trfree-whK}
{\widehat K}_{kl} {\widehat K}^{kl}= \interior{{\widehat K}}_{kl} \interior{{\widehat K}}^{kl} + \tfrac12\,({\widehat K}{}^l{}_{l})^2\,.
\end{equation}
In virtue of these simple observations \eqref{propW} reads as
\begin{align}\label{propWnew}
\mathscr{L}_{\rho} W = {}& -\tfrac12\int_{\mathscr{S}_\rho} \big({\widehat{N}} {\widehat K}{}^l{}_{l}\big)\left[\, 2\,{\widehat R} - ({\widehat K}{}^l{}_{l})^2 \,\right]{\widehat{\boldsymbol{\epsilon}}}\, 
\nonumber \\ {}& + \int_{\mathscr{S}_\rho} \big({\widehat{N}}{\widehat K}{}^l{}_{l}\big)\left[\, {}^{{}^{(3)}}\hskip-1mm R + \interior{{\widehat K}}{}_{kl} \interior{{\widehat K}}{}^{kl} +   2\,{\widehat N}^{-2}\,{\widehat \gamma}^{kl}\,({\widehat D}{}_k {\widehat N}) ({\widehat D}_l {\widehat N}) \,\right]{\widehat{\boldsymbol{\epsilon}}} \,, 
\end{align}
where the vanishing of the integral of the total divergence ${\widehat D}{}^l ({\widehat N}^{-1}\!{\widehat D}_l {\widehat N} )$ was also used. 

\medskip

Note that if in \eqref{propWnew} the factor ${\widehat{N}} {\widehat K}{}^l{}_{l}$ could be replaced by its average 
\begin{equation}\label{eq: average}
\overline{{\widehat{N}} {\widehat K}{}^l{}_{l}}=\frac
{\int_{\mathscr{S}_\rho} {\widehat{N}} {\widehat K}{}^l{}_{l}\,\,{\widehat{\boldsymbol{\epsilon}}}}{\int_{\mathscr{S}_\rho} {\widehat{\boldsymbol{\epsilon}}}}
\end{equation}
the variation of $W$ would simplify considerably. Recall that the integrals in \eqref{eq: average} had already been applied in \eqref{eq: varA}, and it is immediate to see that 
\begin{equation}
\overline{{\widehat{N}} {\widehat K}{}^l{}_{l}}= \mathscr{L}_{\rho} (\log\mathscr{A}_{\rho})\,.
\end{equation}

\medskip

Accordingly, if the product ${\widehat{N}} {\widehat K}{}^l{}_{l}$ could be replaced by its average---i.e.~if the constancy of 
${\widehat{N}} {\widehat K}{}^l{}_{l}$ could be guaranteed  on the individual $\rho=const$ level sets---then \eqref{eq: gen-var-GE} and \eqref{propWnew} would allow us to conclude that
\begin{align}\label{eq: propWnewRed}
\mathscr{L}_{\rho} \,{M}_{\mathcal{G}}  {}& = \frac{\mathscr{A}_\rho^{1/2}}{64 \pi^{3/2}}\, \left[\,\mathscr{L}_{\rho} W + \tfrac12\,\mathscr{L}_{\rho} (\log\mathscr{A}_{\rho})\cdot W \,\right] \nonumber\\ {}&  = \frac{1}{16\pi} \,\mathscr{L}_{\rho} \left[\left(\frac{\mathscr{A}_\rho}{4 \pi}\right)^{1/2}\right]\, \int_{\mathscr{S}_\rho} \left[\, {}^{{}^{(3)}}\hskip-1mm R + \interior{{\widehat K}}{}_{kl} \interior{{\widehat K}}{}^{kl} +   2\,{\widehat N}^{-2}\,{\widehat \gamma}^{kl}\,({\widehat D}{}_k {\widehat N}) ({\widehat D}_l {\widehat N}) \,\right]{\widehat{\boldsymbol{\epsilon}}} \,. 
\end{align}

\medskip

What has been established in the foregoing can be summarized by the following.

\begin{proposition}\label{prop: conditions}
	Consider a Riemannian three-space $(\Sigma,h_{ij})$ such that $\Sigma$ satisfies the Condition specified in section \ref{sec: prelim}. Assume that, apart from origins, $\rho: \Sigma\rightarrow \mathbb{R}$ is a smooth radial function such that the $\mathscr{S}_\rho$ level sets, apart from origins, are topological two-spheres. 
	Assume also that a flow $\rho^i = \widehat{N}\,\widehat{n}^i + \widehat{N}^i$ can be chosen on $\Sigma$, and that the relation 
	\begin{equation}\label{eq: prop1}
	{\widehat{N}} {\widehat K}{}^l{}_{l}=\overline{{\widehat{N}} {\widehat K}{}^l{}_{l}}=\mathscr{L}_{\rho} (\log\mathscr{A}_{\rho})
	\end{equation}
	holds at a $\rho=const$ level set.  Then, 
	\begin{itemize}
		
		\item[(i)] the generic variation $\mathscr{L}_{\rho} \,{M}_{\mathcal{G}}$ of the Geroch mass, at $\mathscr{S}_\rho$, is given by \eqref{eq: propWnewRed}, and   
	\end{itemize}

	\begin{itemize}	 
		\item[(ii)] if, in addition, the area ${\mathscr{A}_\rho}$ is non-decreasing, with respect to the flow $\rho^i$, and the inequality
		\begin{equation}\label{eq: prop2}
		\int_{\mathscr{S}_\rho}\Big[{}^{{}^{(3)}}\hskip-1mm R  +  \interior{{\widehat K}}{}_{kl} \interior{{\widehat K}}{}^{kl} + 2\,{\widehat N}^{-2}\,{\widehat \gamma}^{kl}\,({\widehat D}{}_k {\widehat N}) ({\widehat D}_l {\widehat N})\Big]\,{\widehat{\boldsymbol{\epsilon}}} \geq 0 
		\end{equation}
		holds, then the Geroch mass is non-decreasing at ${\mathscr{S}_\rho}$.
	\end{itemize}
\end{proposition}

Notice that even \eqref{eq: prop2} would allow ${}^{{}^{(3)}}\hskip-1mm R$ somewhere or, when either $\interior{{\widehat K}}{}_{kl}$ or ${\widehat D}{}_k {\widehat N}$ does not vanish, everywhere to be slightly negative on $\mathscr{S}_\rho$. Nevertheless, as in most of the conventional constructions $\interior{{\widehat K}}{}_{kl}$ and $\widehat N$ get to be known only at the very end, usually ${}^{{}^{(3)}}\hskip-1mm R\geq 0$ is assumed in the pertinent arguments \cite{HusIlm97, HusIlm01, Bray01, Bray09}. If one would like to use the weakest possible restrictions, it is rewarding to keep in mind that, if the area is non-decreasing, it suffices to require \eqref{eq: prop2} to guarantee that the Geroch mass is non-decreasing at ${\mathscr{S}_\rho}$.

\medskip

Having these observations the main dilemma we have to face originates from the rigidity of the setup we started with. Namely, if both the Riemannian metric $h_{ij}$ and the foliation are fixed then so are the 
mean curvature ${\widehat K}{}^l{}_{l} = {{\widehat \gamma}^{kl}} (D_k\widehat n_l) = D_k{\widehat n}{}^k$ and the lapse ${\widehat{N}}=({\widehat n}{}^i\partial_i\rho)^{-1}$. Indeed, the only ``freedom'' still remained is nothing but a simple relabeling $\overline{\rho} = \overline{\rho}(\rho)$ of the level sets of the foliation which cannot yield more than the trivial rescaling ${{\widehat{N}}} \rightarrow {\widehat{N}} (\textrm{d}{\rho}/\textrm{d}\overline{\rho})$ of the lapse. 
Accordingly, the factor ${\widehat{N}} {\widehat K}{}^l{}_{l}$ in \eqref{propWnew} is  not constant but, at best, it is merely a smooth function on the $\mathscr{S}_\rho$ level sets of the foliation. 

\subsection{The alternative ways of getting control on monotonicity}\label{subsec: getting-control}

It is rewarding to have a glance again at the structures we have by hand. 
We started with a Riemannian metric $h_{ij}$ defined on a three-surface $\Sigma$ foliated by topological two-spheres. The foliation was fixed by choosing a function $\rho:\Sigma \rightarrow \mathbb{R}$ which, apart from origins, is smooth with a well-defined and non-vanishing gradient $\partial_i\rho$. In addition, a flow $\rho^i$ was also chosen such that $\rho^i\partial_i \rho=1$ holds---apart from origins---everywhere on $\Sigma$. 

Once we have a foliation and a flow local coordinates $(\rho,x^A)$ adapted to $\rho^i$ can be introduced such that $\rho^i=(\partial_\rho)^{i}$, whereas the shift and the induced metric can be given as a two-vector ${\widehat N}{}^A$ and a $2\times 2$ positive definite matrix ${\widehat \gamma}_{AB}$, respectively, such that both smoothly depend on the coordinates $\rho, x^A$. (The capital Latin indices always take the values $2,3$.) In particular, in these coordinates, the line element of the Riemannian metric $h_{ij}$ reads as 
\begin{equation}\label{eq: leh}
{\rm d}s^2 = {\widehat N}{}^2 {\rm d}\rho^2 + {\widehat \gamma}_{AB}\,\big(\,{\rm d}x^A+{\widehat N}{}^A{\rm d}\rho\,\big)\,\big(\,{\rm d}x^B+{\widehat N}{}^B{\rm d}\rho\,\big)\,.
\end{equation}

In summing up, we can say that those Riemannian three-spaces $(\Sigma,h_{ij})$, where $\Sigma$ can be  foliated by topological two-spheres and a flow had been chosen on $\Sigma$, can be 
represented by either of the sets $\{h_{ij}\,; \,\rho: \Sigma \rightarrow \mathbb{R}\,, \rho^i \}$ or $\{{\widehat N},{\widehat N}{}^A, {\widehat \gamma}_{AB}\,; \,\rho: \Sigma \rightarrow \mathbb{R}\,, \rho^i=(\partial_{\rho})^i \}$. If, for instance, it is desirable to have control on the monotonous behavior of the Geroch mass, then, in virtue of Proposition \ref{prop: conditions}, sensible choices for certain maximal subsets have to be made such that the left out ingredients yet have to be constructed in such a way that guarantees \eqref{eq: prop1} and \eqref{eq: prop2} to hold. 
For instance, if the Riemannian metric $h_{ij}$ on $\Sigma$ is preferred to be fixed, then the flow and foliation have to be constructed. This is indeed the path laid down by Geroch in \cite{geroch} by proposing the use of inverse mean curvature flow (see Section\,\ref{sec: IMCF} below).

Alternatively, one may prefer to start with a globally well-defined smooth foliation $\rho: \Sigma \rightarrow \mathbb{R}$ and flow $\rho^i=(\partial_{\rho})^i$ and with a smooth distribution of Riemannian two-metrics, ${\widehat \gamma}_{AB}$, on the $\rho=const$ level sets in $\Sigma$. To construct Riemannian three-spaces such that the $\rho=const$ level sets form a smooth inverse mean curvature foliation, and such that the Geroch mass is non-decreasing while moving outward in the foliation, then the lapse and shift are to be chosen such that both \eqref{eq: prop1} and \eqref{eq: prop2} hold throughout $\Sigma$.

\section{Inverse mean curvature flows and foliations}\label{sec: IMCF}

As indicated above, once the three-metric is fixed, to get control of the monotonous behavior of the Geroch mass, the foliation has to be constructed dynamically. In virtue of Proposition \ref{prop: conditions}, it is not incidental that each of the known attempts aiming to get foliations with non-decreasing Geroch mass \cite{geroch, JangWald77, Jang78, Kijowski-1986, Jezierski-Kijowski-1987, Joerg01, Szabados-2004} essentially starts by specifying a two-sphere and construct the other members of the desired foliation by flowing this initial two-sphere in $\Sigma$ using an {\it inverse mean curvature flow}\,\footnote{Note that there is a much higher variate of dynamically determined foliations and flows to be applied. The so-called $\beta$-foliations proposed by Jacek Jezierski in \cite{Jezierski-1994, Jezierski-1994b, Jezierski-Kijowski-2004}---generalizing the ``conformal harmonic gauge fixing'' (corresponding to $\beta=1$) introduced originally by Jerzy Kijowski \cite{Kijowski-1986} and studied in some details by Piotr Chru\'sciel in \cite{Chrusciel-1986}---are excellent examples of these types. Note, however, that, likewise in the case of the  IMCF, proving the global existence of $\beta$-foliations is also a notoriously difficult problem, and, as far as the author knows, this has not been done yet apart from simple spherically symmetric configurations.} (IMCF). The simplest possible form of such a flow, proposed originally by Geroch in \cite{geroch}, is  
\begin{equation}\label{eq: IMCF0}
\rho^i_{{}_{\{IMCF\}}}=
\big({\widehat K}{}^l{}_{l}\big)^{-1}\,{\widehat{n}}{}^i\,.
\end{equation}
Whenever the global existence of the corresponding foliation can be shown ${\widehat{N}}{\widehat K}{}^l{}_{l} \equiv 1$ holds automatically, and the Geroch mass is non-decreasing with respect to this flow provided that \eqref{eq: prop2} is satisfied. As in this process the foliation $\rho: \Sigma \rightarrow \mathbb{R}$ and, in turn, the lapse ${\widehat N}=({\widehat n}{}^i\partial_i\rho)^{-1}$ and $\interior{{\widehat K}}{}_{kl}$, get to be known only at the very end of the construction the inequality ${}^{{}^{(3)}}\hskip-1mm R \geq  0$ is imposed (see, e.g.~\cite{geroch,HusIlm97,HusIlm01}) to guarantee \eqref{eq: prop2} to hold. 

\medskip

In virtue of the observations made in the previous subsections this construction can also be carried out by using slightly more general form of flows. First, by applying a relabeling $\overline{\rho}=\overline{\rho}(\rho)$ such that ${\widehat{N}}{\widehat K}{}^l{}_{l} \equiv 1$ can be replaced by the relation ${\widehat{N}} {\widehat K}{}^l{}_{l}=\overline{{\widehat{N}} {\widehat K}{}^l{}_{l}}=\mathscr{L}_{\rho} (\log\mathscr{A}_{\rho})$. In addition---however counter intuitive it looks like, especially in virtue of the insensitivity of the variation of the area and the Geroch mass to the shift---we can also add a ``shift part'' to the rescaled IMCF to get the more general flow,\,\footnote{Notably, the $\beta$-foliations introduced by Jezierski \cite{Jezierski-1994,Jezierski-1994b,Jezierski-Kijowski-2004}, via setting up ``gauge conditions'' in a completely coordinate dependent way (see, e.g.~eqs.~(3) and (6) in \cite{Jezierski-1994}), when $\beta=0$ and when an inverse mean curvature foliation labeled by the area radius is applied---neither of the later two conditions was mentioned in \cite{Jezierski-1994,Jezierski-1994b,Jezierski-Kijowski-2004}---can be seen to reproduce a restricted class of these flows.}
\begin{equation}\label{eq: GIMCF}
\rho^i
=
\mathscr{L}_{\rho} (\log\mathscr{A}_{\rho})\cdot({\widehat K}{}^l{}_{l}\big)^{-1}\,{\widehat{n}}{}^i+{\widehat{N}}{}^i\,.
\end{equation}
It is important to keep in mind that, by virtue of \eqref{hatextcurv} and \eqref{rhovf}, ${{\widehat{N}} {\widehat K}{}^l{}_{l}}$ and ${\widehat{N}}{}^i$ are related via
\begin{equation}\label{eq: shift0}
	{{\widehat{N}} {\widehat K}{}^l{}_{l}} = \tfrac12\,{\widehat \gamma}^{ij}\mathscr{L}_{\rho} {\widehat \gamma}_{ij} - \widehat{D}{}_i\widehat{N}{}^i
\end{equation}
or---whenever local coordinates adapted to the foliation and the flow are applied---via
\begin{equation}\label{eq: shift1}
	\widehat{D}{}_A\widehat{N}{}^A= \mathscr{L}_{\rho} \log\big[\sqrt{\det({\widehat \gamma}_{AB})}\big]   - \mathscr{L}_{\rho} (\log\mathscr{A}_{\rho})\,.
\end{equation} 
In subsection \ref{subsec: dyn_shift} we shall return to the solvability of this equation for ${\widehat{N}}{}^A$ provided that ${\widehat \gamma}_{AB}$ is given on the $\mathscr{S}_\rho$ level sets. Note also that if $\widehat{N}{}^A$ was given \eqref{eq: shift1} would yield, instead, a restriction on ${\widehat \gamma}_{AB}$. 

\medskip

The flow given by \eqref{eq: GIMCF} could also be used, in practice, as follows. 
Start by choosing a mean-convex topological two-sphere ${\mathscr{S}}$ in $\Sigma$, and an arbitrary but small positive real number $A>0$ and set the initial value, ${}_{(0)}\widehat{N}$, 
to be the positive function ${}_{(0)}\widehat{N}= A\cdot( {\widehat K}{}^l{}_{l})^{-1}$ on ${\mathscr{S}}$.
Construct now an infinitesimally close two-surface ${\mathscr{S}}'$ simply by Lie dragging the points of ${\mathscr{S}}$ along the auxiliary flow $\rho^i = \widehat{N}\, {\widehat n}{}^i$ in $\Sigma$. By comparing the metric induced on ${\mathscr{S}}$ and ${\mathscr{S}}'$, respectively,  
both terms on the r.h.s.~of \eqref{eq: shift1} can be evaluated on ${\mathscr{S}}'$. In performing the succeeding steps we have to update both the lapse and the shift such that the relation ${\widehat{N}} {\widehat K}{}^l{}_{l}=\overline{{\widehat{N}} {\widehat K}{}^l{}_{l}}$ gets to be maintained in each of these steps. In doing so update first the lapse on ${\mathscr{S}}'$ by setting ${\widehat{N}} =\mathscr{L}_{\rho} (\log\mathscr{A}_{\rho})\cdot ({\widehat K}{}^l{}_{l})^{-1}$, where $\mathscr{L}_{\rho} (\log\mathscr{A}_{\rho})$ is the positive real number determined via the infinitesimal step just had made. The key point here is that one can also update the shift on ${\mathscr{S}}'$---such that ${\widehat{N}} {\widehat K}{}^l{}_{l}=\overline{{\widehat{N}} {\widehat K}{}^l{}_{l}}$ holds there---simply by solving \eqref{eq: shift1} for $\widehat{N}{}^A$ as shown in subsection \ref{subsec: dyn_shift}.  

The succeeding infinitesimal step can then be performed by Lie dragging the points of ${\mathscr{S}}'$ along the flow $\rho^i = \widehat{N}\, {\widehat n}{}^i+{\widehat N}{}^i$ with lapse and shift determined on ${\mathscr{S}}'$ as indicated above. This way, we get the next (infinitesimally close) two-surface  ${\mathscr{S}}''$. By performing analogous infinitesimal steps ultimately, we get a one-parameter family of two-surfaces ${\mathscr{S}}_{\rho}$ foliating a one-sided neighborhood of ${\mathscr{S}}$ in $\Sigma$ such that the product  ${\widehat{N}} {\widehat K}{}^l{}_{l}$ is guaranteed to be a positive constant on each of the individual level sets. 

\medskip

It is important to emphasize that the above-outlined construction by no means is proving the existence of an inverse mean curvature flow. Even a local existence proof requires the use of a suitable parabolic equation (for more details, see \cite{HusIlm01, HusIlm97, Bray01, Bray09}). 

It is also rewarding to keep in mind that
the vanishing of ${\widehat K}{}^l{}_{l}$, which in a non-singular setup corresponds to the vanishing of $\widehat{N}{}^{-1}\cdot\mathscr{L}_{\rho} (\log\mathscr{A}_{\rho})$, could get in the way of applicability of the flow in \eqref{eq: GIMCF}. In particular, as ${\widehat K}{}^l{}_{l}$ vanishes at minimal and maximal surfaces, they do represent naturally limits to the domains in $\Sigma$, where the above-outlined construction can be applied. Note, however, that the occurrence of minimal and maximal surfaces depends on the choice we make for a timeslice in the ambient space. For instance, while the bifurcation surface of the  Schwarzschild spacetime is a minimal surface on the standard Schwarzschild $t_{Schw}=const$ timeslices,  the Kerr-Schild $t_{KS}=const$ timeslices of the same spacetime can be foliated by metric spheres with area radius ranging from zero to infinity such that neither of the $r=const$ level sets is extremal. It is also important to keep in mind that the use of the flow in \eqref{eq: GIMCF} does not require $\Sigma$ to be `time symmetric' or maximal. In particular, we nowhere required the three-scalar curvature, ${}^{{}^{(3)}}\hskip-1mm R$, to be non-negative. 

\medskip

Note also that the most serious issue, namely, the global existence and regularity of foliations yielded by an inverse mean curvature flow\,\footnote{The level of the involved technicalities gets to be transparent in the proof of the Riemannian Penrose inequality by Huisken and Ilmanen \cite{HusIlm97,HusIlm01}, or in that of the corresponding higher dimensional generalization by Bray \cite{Bray01}, Bray and Lee \cite{Bray09}.} does not get to be relaxed by applying the flow \eqref{eq: GIMCF}. To see this recall that the level sets of foliation constructed by \eqref{eq: IMCF0} remain intact while they get to be relabeled and combined by the integral curves of the flow determined by \eqref{eq: GIMCF}.  In virtue of these observations the introduction of this more general looking inverse mean curvature flow may appear to be completely superfluous. The rest of this paper is to convince the readers that things are in order. More precisely, it is shown that by applying the flexibility provided by involving flows with a non-trivial shift, a large variety of Riemannian three-spaces can be constructed such that each will be endowed with a smooth inverse mean curvature foliation and with some additional desirable properties.

\section{The new construction}\label{sec: new-construction}

This section describes the construction that allows producing an inverse mean curvature foliation from a given smooth foliation. We start with a one-parameter family of Riemannian two-metric $\widehat{\gamma}_{AB}$ given on the $\mathscr{S}_\rho$ level sets of a foliation of the three-manifold $\Sigma$ satisfying the Condition specified in section \ref{sec: prelim}. The $\mathscr{S}_\rho$ level sets are topological two-spheres which are determined by a function $\rho:\Sigma\rightarrow \mathbb{R}$ that, apart from origins, is smooth. Accordingly, it is assumed that the gradient $\partial_i\rho$ is well-defined and non-vanishing everywhere, apart from origins, and that a flow $\rho^i$ had also been chosen such that $\rho^i\partial_i\rho=1$, apart from these origins, throughout $\Sigma$. Our main task is to
show that we can always choose the shift such that the $\mathscr{S}_\rho$ level sets constitute a smooth inverse mean curvature foliation of $\Sigma$ for the three-metric constructed out of suitable data ${\widehat N}, {\widehat N}{}^A$ and $\widehat{\gamma}_{AB}$ via \eqref{eq: leh}. 

\subsection{Determining the shift}\label{subsec: dyn_shift}

In proceeding first we show that---while treating the foliation $\rho: \Sigma \rightarrow \mathbb{R}$, the flow $\rho^i=(\partial_{\rho})^i$ and the metric ${\widehat \gamma}_{AB}$, on the $\mathscr{S}_\rho$ level sets, as prescribed fields on $\Sigma$---equation \eqref{eq: shift1} can always be solved for the shift.

\medskip

Before solving \eqref{eq: shift1} it is important to make the following consistency check. Clearly, the integral of both sides of \eqref{eq: shift1}, when evaluated on any of the ${\mathscr{S}}_\rho$ level sets, must vanish. The integral of the total divergence on the l.h.s.~is obviously zero, whereas the integral of the r.h.s.~can also be seen to vanish by virtue of the relations 
\begin{equation}
\int_{\mathscr{S}_\rho} \hskip-0.2cm\mathscr{L}_{\rho} \log\Big[\sqrt{\det({\widehat \gamma}_{AB})}\Big] \, {\widehat{\boldsymbol{\epsilon}}}  = \int_{\mathscr{S}_\rho} \mathscr{L}_{\rho} \Big[\sqrt{\det({\widehat \gamma}_{AB})} \Big]\, {{\boldsymbol{\varepsilon}}} = \mathscr{L}_{\rho} \Big[\int_{\mathscr{S}_\rho} \hskip-0.2cm {\widehat{\boldsymbol{\epsilon}}} \,\Big]=\mathscr{L}_{\rho}(\log \mathscr{A}_{\rho})\!\int_{\mathscr{S}_\rho} \hskip-0.2cm {\widehat{\boldsymbol{\epsilon}}} \,,
\end{equation} 
where ${{\boldsymbol{\varepsilon}}}={\widehat{\boldsymbol{\epsilon}}}/\sqrt{\det({\widehat \gamma}_{AB})}$, along with its $\rho$-invariance, was applied. 

Notably, solving \eqref{eq: shift1} is easier than it appears at first glance. To see this, recall that as the first Betti number of topological two-spheres is zero, they admit only the trivial harmonic form. This allows us, by making use of the Hodge decomposition, to represent the shift vector, on any of the $\mathscr{S}_{\rho}$ level sets, via a pair of smooth potentials $\chi$ and $\eta$ as
\begin{equation}\label{eq: hodge}
\widehat{N}{}^A=\widehat{D}{}^{A}\chi+\widehat{\epsilon}{}^{AB}\widehat{D}{}_{B}\eta \,.
\end{equation}
The first and the second terms on the r.h.s. of \eqref{eq: hodge} are the longitudinal and transversal parts of ${\widehat N}{}^A$, respectively.
Notably \eqref{eq: shift1} can then be seen to take the form 
\begin{equation}\label{eq: shift_chi}
\widehat{D}{}^A\widehat{D}{}_A \chi= \mathscr{L}_{\rho} \log\big[\,\sqrt{\det({\widehat \gamma}_{AB})}\,\big]   - \mathscr{L}_{\rho} (\log\mathscr{A}_{\rho})
\end{equation}
which is an elliptic equation for $\chi$. Solutions to \eqref{eq: shift_chi} can, in principle, be given in terms of the coefficients of the expansion of the r.h.s.~of \eqref{eq: shift_chi} with respect to the eigenfunction of the Laplacian $\widehat{D}{}^A\widehat{D}{}_A$ \cite{Chavel-1984}. 
As all the coefficients and source terms in this elliptic equation smoothly depend on the one-parameter family of smooth metrics, ${\widehat \gamma}_{AB}$, solutions to \eqref{eq: shift_chi} do exist---they are unique up to the ``monopole'' part--- and they do also smoothly depend on $\rho$.

It is an important to know whether the above-outlined determination of the shift is compatible with the middle relation of \eqref{eq: origin} when considerations are restricted to regular origins. In verifying that this is so the aim is to show that the proposed procedure does yield a shift vector field that satisfies the middle relation of \eqref{eq: origin}, in area-radial coordinate $\rho$ defined in a neighborhood of a regular origin provided that only the third relation of \eqref{eq: origin} restricting the two-metric there is allowed to be applied. In proceeding notice first that, in virtue of the third relation of  \eqref{eq: origin}, the source on the right hand side of \eqref{eq: shift_chi} must have the functional form $(\rho-\rho_*)\,\psi_1+\mathcal{O}[(\rho-\rho_*)^2]$, where $\psi_1$  is a smooth function in a neighborhood of the regular origin at $\rho=\rho_*$. This, in virtue of \eqref{eq: shift_chi}, implies then that $\chi$, and, in turn, $\partial_A\chi\sim \mathcal{O}[(\rho-\rho_*)^3]$ or $\partial^A\chi\sim (\rho-\rho_*)^{-2}\,\interior{\gamma}{}^{AB}\partial_B\chi\sim \mathcal{O}[(\rho-\rho_*)^1]$. By choosing then the freely specifiable potential $\eta$ to be the product of a smooth bounded function and $(\rho-\rho_*)^3$ in a neighborhood of the regular origin at $\rho=\rho_*$, in virtue of the third relation of \eqref{eq: origin}, the second term $\widehat{\epsilon}{}^{AB}\partial_{B}\eta\sim ((\rho-\rho_*)^{-2}\,\interior{\gamma}{}^{AE})((\rho-\rho_*)^{-2}\,\interior{\gamma}{}^{BF})((\rho-\rho_*)^{2}\,\interior{\epsilon}{}_{EF})\, \partial_{B}\eta$ is also of order $\mathcal{O}[(\rho-\rho_*)^1]$  in a neighborhood of the regular origin at $\rho=\rho_*$. Combining these simple observations, it is straightforward to see that the shift ${\widehat N}{}^A$, determined via \eqref{eq: hodge}, does also satisfy the middle relation of \eqref{eq: origin}.

Summing up the above observations, we have then the following. 

\begin{theorem}\label{theor: shift}
	Consider a smooth, three-dimensional manifold $\Sigma$ satisfying the Condition specified in section \ref{sec: prelim}. Assume that $\rho:\Sigma\rightarrow \mathbb{R}$ is a function such that, apart from origins, it is smooth, and the $\rho=const$ level sets are topological two-spheres. Assume that a smooth one-parameter family of Riemannian two-metric ${\widehat \gamma}_{AB}$ on the  $\mathscr{S}_\rho$ level sets, along with a flow $\rho^i$ that is smooth, apart from origins, had also been chosen on $\Sigma$. 
	Then, apart from origins, there exists an, up to the monopole part, the unique smooth solution to \eqref{eq: shift_chi} on the $\mathscr{S}_\rho$ level sets such that, regardless of choice made for the potential $\eta$ and the lapse ${\widehat N}$, for the corresponding smooth vector field ${\widehat N}{}^A$ \eqref{eq: shift1} holds, apart from origins, on $\Sigma$.
\end{theorem}

Note that whenever ${\widehat N}{}^A$ is chosen as described above then---irrespective of the choice made for the other potential $\eta$ and for the lapse ${\widehat N}$---the constructed Riemannian three-metric will be such that ${\widehat{N}} {\widehat K}{}^l{}_{l}=\overline{{\widehat{N}} {\widehat K}{}^l{}_{l}}=\mathscr{L}_{\rho} (\log\mathscr{A}_{\rho})$ holds on the individual $\rho=const$ surfaces. In fact, the most important implication of Theorem \ref{theor: shift} is that the topological two-spheres $\mathscr{S}_\rho$ form an inverse mean curvature foliation of $\Sigma$ independent of the choice made for $\eta$ and ${\widehat N}$. In particular, if the lapse is chosen to be constant on the individual $\mathscr{S}_\rho$ level sets then the trace ${\widehat{K}}{}^l{}_l={\widehat{N}}^{-1}\mathscr{L}_{\rho} (\log\mathscr{A}_{\rho})$ gets also to be constant on them. In this special case the $\mathscr{S}_\rho$ level sets do also form a CMC foliation of $\Sigma$. 

\medskip

It is also important to emphasize that the statement of Theorem \ref{theor: shift} also holds in a wider context. Consider a generic smooth, three-dimensional manifold $\Sigma$. As discussed in section\,\ref{sec: prelim}, whenever $\Sigma$ is (almost everywhere) foliated by topological two-spheres, then  $\Sigma$ can be given as the disjoint union of disks and cylinders that are glued together via $\rho=const$ slices through index one or two non-degenerate critical points of a suitable Morse function $\rho: \Sigma\rightarrow \mathbb{R}$. Assume that a smooth distribution of two-metric ${\widehat \gamma}_{AB}$ can be introduced on the $\rho=const$ level sets of $\Sigma$. For instance, smooth distribution of this type can always be given by starting with a smooth auxiliary Riemannian three-metric ${\widetilde h}_{ij}$ on $\Sigma$ and by determining the two-metric induced by ${\widetilde h}_{ij}$ on each of the $\mathscr{S}_{\rho}$ level sets. 
Recall now that the determination of the $\chi$-potential, and in turn of the shift ${\widehat N}{}^A$, as described above, can be carried out level set by level set, which by smoothness of all the geometric ingredients yields an inverse mean curvature foliation on $\Sigma$, that is smooth everywhere apart from the $\rho=const$ slices through the non-degenerate critical points. Note that then, in virtue of \eqref{eq: prop1}, the function $\partial_\rho(\log\mathscr{A}_{\rho})$ is constant on any of those regular $\mathscr{S}_\rho$ level set. Consider now a critical slice $\rho=\overline{\rho}$ and an arbitrary point $p$ on this level set. By smoothness of the Morse function  $\rho: \Sigma\rightarrow \mathbb{R}$ the point $p$, as any other point on the critical slice $\mathscr{S}_{\overline{\rho}}$, can always be represented as an accumulation point of a  point sequence $\{p_i\}$, with $p_i \in\mathscr{S}_{\rho_i}$, such that $\{\rho_i\} \rightarrow \overline{\rho}$. As the function $\mathscr{L}_{\rho} (\log\mathscr{A}_{\rho})$ is constant on each of the $\mathscr{S}_{\rho_i}$ regular level sets for any choice of  $\{p_i\}$  the sequence $\{\mathscr{L}_{\rho} (\log\mathscr{A}_{\rho})\vert_{p_i}\}$ of real numbers by construction must tend to a common limit value, denoted by $\mathscr{L}_{\rho} (\log\mathscr{A}_{\overline{\rho}})$. This, in turn, guarantee then that $\mathscr{L}_{\rho} (\log\mathscr{A}_{\rho})$ is a smooth function on $\Sigma$ such that it is constant on each of the $\rho=const$ level sets.

The aforementioned smoothness properties do also guarantee that the potentials $\chi$ and $\eta$, and, as well as the shift, determined by them via \eqref{eq: hodge}, extend smoothly to the critical slices, and, in turn, also to the entire of $\Sigma$. 

The following is a summing up of what has just been verified.

\begin{corollary}\label{corolary1}
	Consider a smooth, three-dimensional manifold $\Sigma$. Assume that $\Sigma$ is generic, i.e., there exists a (smooth) Morse function $\rho:\Sigma\rightarrow \mathbb{R}$ such that, apart from the $\rho=const$ slices through the isolated critical points, the connected components of the $\rho=const$ level sets are topological two-spheres. Assume that a smooth distribution of Riemannian two-metrics ${\widehat \gamma}_{AB}$ on the $\mathscr{S}_\rho$ level sets, along with a smooth flow $\rho^i$ had also been chosen on $\Sigma$. Then, the function $\mathscr{L}_{\rho} (\log\mathscr{A}_{\rho})$ is smooth throughout $\Sigma$ such that it is constant on the individual $\rho=const$ level sets. Besides, there exists a shift vector field such that, apart from critical points, ${\widehat N}{}^A$ is smooth, and it satisfies \eqref{eq: shift1} on $\Sigma$.
\end{corollary} 

\subsection{Choosing the lapse}\label{subsec: dyn_lapse}

In this subsection, first conditions on the lapse and the area will be identified that guarantee the non-negativity of the Geroch mass on the individual $\mathscr{S}_\rho$ level sets. An integrodifferential expression is also derived that can be used to characterize the $\rho$-dependence of the Geroch mass there. 

\medskip

Assuming that the shift ${\widehat N}{}^A$ is chosen as in Theorem \ref{theor: shift}, i.e.~the relation ${\widehat N}{{\widehat K}^l}{}_{l} =\overline{{\widehat N}{{\widehat K}^l}{}_{l}} =\mathscr{L}_{\rho} (\log\mathscr{A}_{\rho})$ holds on each of the $\mathscr{S}_\rho$ level sets, the Geroch mass can be rephrased, in terms of the area and lapse, as 
\begin{align}\label{eq: GQLM2}
{M}_{\mathcal{G}} = {} & \frac{\mathscr{A}_\rho^{1/2}}{64 \pi^{3/2}}\int_{\mathscr{S}_\rho}\!\! \Big[\, 2\,{\widehat R} - ({\widehat K}{}^l{}_{l})^2 \,\Big]\,{\widehat{\boldsymbol{\epsilon}}} =\frac{\mathscr{A}_\rho^{1/2}}{64 \pi^{3/2}}\,\Big[\,16\pi -\! \int_{\mathscr{S}_\rho}\![\,\mathscr{L}_{\rho} (\log\mathscr{A}_{\rho})\,]^2\,{\widehat N}{}^{-2} \,{\widehat{\boldsymbol{\epsilon}}} \,\Big] \nonumber \\ = {} &
\frac{\mathscr{A}_\rho^{1/2}}{64 \,\pi^{3/2}} \,\Big[\,16\pi - \int_{\mathscr{S}_\rho}[\,\mathscr{L}_{\rho} (\log\mathscr{A}_{\rho})\,]^2{\widehat N}{}^{-2}\, \Big(\frac{\mathscr{A}_{\rho}}{4\pi}\Big) \,\widetilde{\boldsymbol{\epsilon}}\,\Big] \nonumber \\ = {} & \frac{1}{8\pi}\Big(\frac{\mathscr{A}_\rho}{4\pi}\Big)^{1/2}\!\!
\!\int_{\mathscr{S}_\rho}\!\Big[\,1-\Big(\,\mathscr{L}_{\rho} \Big(\frac{\mathscr{A}_\rho}{4\pi}\Big)^{1/2}\,\Big)^2\widehat N{}^{-2}\Big]\,\widetilde{\boldsymbol{\epsilon}}
\,,
\end{align}
where in the last but one step the ``normalized volume element'' $\widetilde{\boldsymbol{\epsilon}}=\Big({\mathscr{A}_\rho}/{(4\pi)}\Big)^{-1}\,{\widehat{\boldsymbol{\epsilon}}}$, satisfying the relation $\int_{\mathscr{S}_\rho} {\widetilde{\boldsymbol{\epsilon}}}=4\pi$ was introduced.

\medskip

In many cases it is advantageous to use the ``area-radial coordinate'', i.e., to choose $\rho: \Sigma \rightarrow \mathbb{R}$ such that it satisfies the relation $\mathscr{A}_\rho = 4\pi\rho^2$. Nevertheless, in advance of applying, it is rewarding to have a glance again at the relation
\begin{equation}\label{eq: area-radial}
{{\widehat K}^l}{}_{l} = {\widehat N}^{-1}\mathscr{L}_{\rho} (\log\mathscr{A}_{\rho})\,.
\end{equation}
Recall that \eqref{eq: area-radial} is guaranteed to hold---for any choice of the $\rho$-parameter---as far as the $\mathscr{S}_{\rho}$ level sets do form an inverse mean curvature foliation of $\Sigma$. In virtue of \eqref{eq: area-radial} if for some value of $\rho$ the $\mathscr{S}_{\rho}$ level set is extremal, i.e.~${{\widehat K}^l}{}_{l} =0$ then either ${\widehat N}^{-1}$ or $\mathscr{L}_{\rho} (\log\mathscr{A}_{\rho})$ has to vanish. Accordingly, if one uses area-radial coordinate ${\widehat N}$ cannot be bounded on such an extremal surface as $\mathscr{A}_\rho = 4\pi\rho^2$ implies that $\mathscr{L}_{\rho} (\log\mathscr{A}_{\rho})=2/\rho$ which does not vanish. Similarly, if ${\widehat N}$ is demanded to be smooth and bounded everywhere on $\Sigma$ then either there is no extremal $\mathscr{S}_{\rho}$ level set on $\Sigma$ or, alternatively, $\rho$ cannot be an area-radius throughout $\Sigma$. Even if there are extremal surfaces on  $\Sigma$ if they are isolated then area-radial coordinate can be introduced between any of the succeeding pairs of them\,\footnote{However extreme it sounds, there may exist $\{\rho_i\}$ sequences possessing an accumulation point ${\overline \rho}$ such that ${{\widehat K}^l}{}_{l}$ vanishes at each of the $\rho={\rho_i}$ level sets.  Special care is required then to investigate what happens at the accumulation surface $\mathscr{S}_{\overline\rho}$ though ${{\widehat K}^l}{}_{l}$ must also vanish there. Clearly, the use of area-radial coordinate in its neighborhood appears to be completely adverse, nevertheless, the author is indebted to an unknown referee for pointing out the possible occurrence of such cases.}. In addition, it may also happen that ${{\widehat K}^l}{}_{l}$ vanishes on some tubular subsets of $\Sigma$. Note, finally, that independent of the above discussed particular cases, in virtue of \eqref{eq: GQLM}, the relation 
\begin{equation}\label{eq: Penrose-loc0}
\mathscr{A}_{\rho} = 16\pi \,[{M}_{\mathcal{G}}(\rho)] ^{\,\,2} 
\end{equation} 
must hold on any of the extremal $\mathscr{S}_{\rho}$ level sets. 

\medskip

Once the area-radial coordinate $\rho$ can rightfully be applied by substituting the relations $\mathscr{A}_\rho = 4\pi\rho^2$ and $\mathscr{L}_{\rho} (\log\mathscr{A}_{\rho})={2}/{\rho}$ into \eqref{eq: GQLM2}, the Geroch mass simplify to 
\begin{align}\label{eq: GQLM22}
{M}_{\mathcal{G}} = \frac{1}{8\pi}\, 
\!\int_{\mathscr{S}_\rho}\!\rho\,\Big(\,1-\widehat N{}^{-2}\Big)\,\widetilde{\boldsymbol{\epsilon}}
\,.
\end{align}

\medskip

Notably, in case of quasi-spherical foliations, and only in that special case, the normalized volume element ${\widetilde{\boldsymbol{\epsilon}}}$ reduces to the unit sphere volume element ${\interior{\boldsymbol{\epsilon}}}$. The concept of quasi-spherical foliations---with ${\widehat \gamma}_{AB}=\rho^2\,{\interior \gamma}_{AB}$, where ${\interior \gamma}_{AB}$ stand for the unit sphere metric---was introduced by Bartnik in \cite{Bartnik}. He also introduced a mass aspect function which, in the present notation, reads as $m=\tfrac{1}{2}\,\rho\,\big[\,1 -  {\widehat N}{}^{-2}\,\big]$. Bartnik applied this function in \cite{Bartnik} to verify the global existence of solutions to a parabolic equation that arises in the context of quasi-spherical foliations for the lapse. Indeed, then the quasi-local mass introduced by Bartnik, see equation (1.5) of \cite{Bartnik}, that is an integral of the aforementioned auxiliary mass aspect function, is a special case of the Geroch mass \eqref{eq: GQLM} which takes the form  \eqref{eq: GQLM22} when area-radial coordinates can be applied. 

\medskip

In returning to the generic case, note that in virtue of \eqref{eq: GQLM2} ${\widehat N}$ can always be chosen such that the quasi-local Geroch mass is non-negative throughout $\Sigma$. This verifies then the following. 

\begin{theorem}
	Assume that the conditions of Theorem \ref{theor: shift} hold, i.e.~the shift is such that ${\widehat N}{{\widehat K}^l}{}_{l}=\overline{{\widehat N}{{\widehat K}^l}{}_{l}} =\mathscr{L}_{\rho} (\log\mathscr{A}_{\rho})$ throughout, and also that $\Sigma$ foliated by a one-parameter family of topological two-spheres $\mathscr{S}_\rho$ that are level surfaces of a function $\rho: \Sigma\rightarrow \mathbb{R}$ that is, apart from origins, smooth and the $\partial_i\rho$ gradient of which is well-defined and does not vanish apart from origins. Then, 
\begin{itemize}
	\item[(i)] 	
	the Geroch mass is non-negative on a specific $\rho=const$ level set if the inequality
	\begin{equation}\label{eq: int-MG}
	\int_{\mathscr{S}_\rho}\!\Big[\,1-\Big(\,\mathscr{L}_{\rho} \Big(\frac{\mathscr{A}_\rho}{4\pi}\Big)^{1/2}\,\Big)^2\widehat N{}^{-2}\Big]\,\widetilde{\boldsymbol{\epsilon}} \geq 0
	\end{equation}
	holds, and 
	\item[(ii)] if $\rho$ is the area-radial coordinate, in virtue of \eqref{eq: GQLM22}, regardless of the specific functional dependence of the lapse ${\widehat N}$, the Geroch mass is non-negative on $\Sigma$ provided that 
	\begin{align}\label{eq: GQLM2-cond}
	\int_{\mathscr{S}_\rho}\!\Big(\,1-\widehat N{}^{-2}\Big)\,\widetilde{\boldsymbol{\epsilon}} \geq 0
	\,. 
	\end{align}
	In particular, ${M}_{\mathcal{G}} \geq 0$ on those $\mathscr{S}_\rho$ level sets where ${\widehat N} \geq 1$.
\end{itemize}
	
\end{theorem}

Note that the primary role of the conditions in Theorem \ref{theor: shift} is to guarantee that---for any of the constructed Riemannian three-spaces, the $\mathscr{S}_\rho$ level sets form an inverse mean curvature foliation in $\Sigma$. Consider again---as in the discussion preceding Corollary \ref{corolary1}---a generic smooth, three-dimensional manifold. In virtue of Corollary \ref{corolary1}, by choosing the shift  ${\widehat N}{}^A$ properly, the $\rho=const$ level surface of a Morse function $\rho: \Sigma\rightarrow \mathbb{R}$ are guaranteed to form an inverse mean curvature foliation. This, by appealing to the smoothness of $\Sigma$, along with that of ${\widehat \gamma}_{AB},{\widehat N}{}^A$ and ${\widehat N}$, implies the non-negativity of the Geroch mass on the critical level sets provided that \eqref{eq: int-MG} holds on each of the non-critical slices. 

This verifies then the following. 

\begin{corollary}\label{corolary2}
	Consider a smooth, three-dimensional manifold $\Sigma$. Assume that $\Sigma$ is generic, i.e., there exists a (smooth) Morse function $\rho:\Sigma\rightarrow \mathbb{R}$ such that, apart from the $\rho=const$ slices through the isolated critical points, the connected components of the $\rho=const$ level sets are topological two-spheres. Assume that a smooth distribution of Riemannian two-metrics ${\widehat \gamma}_{AB}$ on the $\mathscr{S}_\rho$ level sets, along with a smooth flow $\rho^i$ had also been chosen on $\Sigma$. Then, by choosing the shift, ${\widehat N}{}^A$ as in Corollary \ref{corolary1}, 
	the Geroch is guaranteed to be non-negative on each of the $\rho=const$ level sets in $\Sigma$ if \eqref{eq: int-MG} holds on them.
\end{corollary} 

\medskip

Recall that in virtue of the argument in section\,\ref{sec: var-Geroch-energy} Geroch mass, with respect to the flow $\rho^i$, is determined by \eqref{eq: propWnewRed}, provided that the ${\mathscr{S}_\rho}$ level sets form an inverse mean curvature foliation. 

All these can be used to verify the following. 

\begin{theorem}\label{theor: lapse}
Assume that the conditions of Theorem \ref{theor: shift} are satisfied, and thereby the shift ${\widehat N}{}^A$ is constructed such that the relation ${\widehat N}{{\widehat K}^l}{}_{l}=\overline{{\widehat N}{{\widehat K}^l}{}_{l}} =\mathscr{L}_{\rho} (\log\mathscr{A}_{\rho})$  holds, apart from origins, on $\Sigma$.
Assume that $\rho: \Sigma \rightarrow \mathbb{R}$ is the area radial coordinate and a smooth positive and bounded lapse, ${\widehat N}$, has also been chosen on $\Sigma$. 
Then, for any $\rho_1 < \rho_2$ the inequality ${M}_{\mathcal{G}}(\rho_1)\leq {M}_{\mathcal{G}}(\rho_2)$ holds provided that the integral inequality
\begin{equation}\label{eq: prop2-int}
\int_{\rho_1}^{\rho_2}\left(\int_{\mathscr{S}_\rho}\Big[{}^{{}^{(3)}}\hskip-1mm R  +  \interior{{\widehat K}}{}_{kl} \interior{{\widehat K}}{}^{kl} + 2\,{\widehat N}^{-2}\,{\widehat \gamma}^{kl}\,({\widehat D}{}_k {\widehat N}) ({\widehat D}_l {\widehat N})\Big]\,{\widehat{\boldsymbol{\epsilon}}} \,\right) \mathrm{d}\rho\geq 0
\end{equation}
is satisfied.

\end{theorem}

\medskip

Notice that \eqref{eq: prop2-int} is less stringent than \eqref{eq: prop2} as it allows the integral in the round bracket to be negative on certain subintervals between $\rho_1$ and  $\rho_2$.  

\subsection{Asymptotically flat configurations}\label{subsec: more results}

In this subsection attention will be restricted to asymptotically flat configurations. In proceeding recall first that a three-metric $h_{ij}$ is asymptotically flat if in the asymptotic region---that is supposed to be diffeormorphic to $\mathbb{R}^3\setminus \mathcal{B}(0,r)$, where $\mathcal{B}(0,r)$ is a ball of radius $r$ centered at the origin in $\mathbb{R}^3$---it approaches the Euclidean metric not slower than $\rho^{-1}$, where $\rho$ stands for the area-radial coordinate in the asymptotic region. In virtue of the results in subsection 2.2.1 of \cite{CsK-IR-2019}, asymptotic flatness of a three-metric is guaranteed if the fields $\widehat  N, \widehat  N{}^A, \widehat \gamma_{AB}$ fall off as 
\begin{equation}\label{hgfall}
\widehat  N{}-1 \sim  \mathscr{O}(\rho^{-1})\,, \quad  \widehat  N^A \sim  \mathscr{O}(\rho^{-3})\,,  \quad \widehat \gamma_{AB} - \rho^2\,\interior{\gamma}{}_{AB} \sim  \mathscr{O}(\rho^{-1}) \,,
\end{equation}
respectively, where $\interior{\gamma}{}_{AB}$ stands here again for the unit sphere metric. 

As an essential self-consistency check, it is important to show that the middle relation in \eqref{hgfall} is compatible with the determination of the shift as described in detail in subsection \ref{subsec: dyn_shift}. To see that this is indeed the case, note first that, 
whenever the freely specifiable two-metric $\widehat \gamma_{AB}$ is arranged to satisfy the third relation in \eqref{hgfall} then the source on the right-hand-side of \eqref{eq: shift_chi} must have the asymptotic form $\psi_{-3}\,\rho^{-3}+\mathscr{O}[\rho^{-4}]$, where $\psi_{-3}$ is a smooth bounded function in the asymptotic region. This, in virtue of the third relation in \eqref{hgfall}, implies that the $\chi$ potential, and, in turn, its gradient $\partial_A\chi$ is guaranteed to decay as $\rho^{-1}$. This implies then that $\partial^A\chi \sim \rho^{-2}\,\interior{\gamma}{}^{AB}\partial_B\,\chi$ falls off not slower than $\rho^{-3}$. 
Since the other potential $\eta$ is freely specifiable, in virtue of the relation $\widehat{\epsilon}{}^{AB}\partial_{B}\eta\sim (\rho^{-2}\,\interior{\gamma}{}^{AE})(\rho^{-2}\,\interior{\gamma}{}^{BF})(\rho^{2}\,\interior{\epsilon}{}_{EF})\, \partial_{B}\eta$, by choosing $\eta$ to be a bounded smooth function on $\Sigma$ that decays as  $\rho^{-1}$, the shift $\widehat  N^A$---constructed out of $\chi$ and  $\eta$ via \eqref{eq: hodge}---is guaranteed to fall off as $\rho^{-3}$ which completes our consistency check.  

\medskip

Consider now a datum $\widehat  N, \widehat  N{}^A, \widehat \gamma_{AB}$ on $\Sigma$ such that the three-metric determined by them, via \eqref{eq: leh}, is asymptotically flat. Then, in virtue of the third relation in  \eqref{hgfall} the area of the  $\mathscr{S}_{\rho}$ level sets is increasing (at least) in the asymptotic region exterior to a $\rho=\rho_1=const$ level set. Assume that $\Sigma$ is either diffeomorphic to $\mathbb{R}^3$ with a regular origin at $\rho=\rho_0$ such that neither of the $\mathscr{S}_{\rho}$ level sets is a minimal surface on $\Sigma$, or $\Sigma$ is cylindrical with an inner boundary $\mathscr{S}_{\rho_0}$ that is an outermost minimal surface on $\Sigma$. In both cases, the area is increasing throughout $\Sigma$, and area-radial coordinates can be introduced everywhere on $\Sigma$. Assume that this has been done. It is also known that if $\Sigma$ is diffeomorphic to $\mathbb{R}^3$ then the Geroch mass ${M}_{G}$ vanishes at the regular origin, whereas whenever $\Sigma$ is cylindrical  $\mathscr{A}_{\rho_0} = 16\pi \,\big[{M}_{G}(\rho_0)\big]^{\,2}$ holds at the $\mathscr{S}_{\rho_0}$ minimal surface. 

\medskip

Assume, in addition, the inequality 
\begin{equation}\label{eq: prop-int}
\int_{\rho_0}^{\infty}\Big(\int_{\mathscr{S}_\rho}\Big[{}^{{}^{(3)}}\hskip-1mm R  +  \interior{{\widehat K}}{}_{kl} \interior{{\widehat K}}{}^{kl} + 2\,{\widehat N}^{-2}\,{\widehat \gamma}^{kl}\,({\widehat D}{}_k {\widehat N}) ({\widehat D}_l {\widehat N})\Big]\,{\widehat{\boldsymbol{\epsilon}}} \,\Big) \mathrm{d}\rho \geq 0
\end{equation}
holds. Then---in virtue of \eqref{eq: propWnewRed}, along with the fact that in the asymptotically flat case (as shown, e.g.~in \cite{JangWald77}) the Geroch mass tends to the Arnowitt-Deser-Misner (ADM) mass, $M_{ADM}$, in the $\rho \rightarrow \infty$ limit---if $\Sigma$ is diffeomorphic to $\mathbb{R}^3$ 
\begin{equation}
	M_{ADM}\geq 0\,,
\end{equation}
whereas in the cylindrical case 
\begin{equation}\label{eq: loca-Penrose}
\mathscr{A}_{\rho_0} = 16\pi \,\big[{M}_{G}(\rho_0)\big]^{\,2} \leq 16\pi \,\Big[\lim_{\rho\rightarrow \infty}{M}_{G}(\rho)\Big]^{\,2}= 16\pi \,{M}_{ADM}^{\,2} 
\end{equation}
can be seen to hold.
All these observations are summed up in the following.

\begin{theorem}\label{theo: asympt}
	Assume that the conditions of Theorem \ref{theor: shift} hold and also that the Riemannian three-space $(\Sigma,h_{ij})$ constructed out of the data  $\widehat  N, \widehat  N{}^A, \widehat \gamma_{AB}$ is asymptotically flat in the exterior to a $\rho=\rho_1=const$ level set.
	
	\begin{itemize}
		
		\item[(i)] If $\Sigma$ is diffeomorphic to $\mathbb{R}^3$, with a regular origin, such that neither of the $\mathscr{S}_{\rho}$ level sets is minimal on $\Sigma$, and the inequality \eqref{eq: prop-int} holds such that it is strict somewhere on $\Sigma$ then the positive mass theorem holds, i.e., ${M}_{ADM}>0$.
		
		\item[(ii)] If $\Sigma$ is cylindrical and  $\mathscr{S}_{\rho_0}$ is an outermost minimal surface on its inner boundary such that the inequality \eqref{eq: prop-int} holds then the Penrose inequality 
		\begin{equation}\label{eq: global Penrose}
		\mathscr{A}_{\rho_0} \leq 16\pi \,{M}_{ADM}^{\,2} 
		\end{equation}
		is satisfied.
	\end{itemize}
\end{theorem}

Note that the inequality \eqref{eq: prop-int}  is much less stringent than \eqref{eq: prop2} as the integral in the round bracket in \eqref{eq: prop-int} may become negative on various subintervals in $[\rho_0,\infty)$. This happens, for instance, if the Geroch mass, while still being non-negative, is decreasing on the corresponding subintervals.

\section{Final remarks}\label{sec: final-remarks}

A new method was introduced that can be used to construct a high variety of Riemannian three-spaces such that each admits a smooth inverse mean curvature foliation. The construction starts by choosing a smooth one-parameter family of Riemannian two-metric $\widehat{\gamma}_{ij}$ on a sphere $\mathscr{S}$. The desired type of Riemannian three-spaces $(\Sigma,h_{ij})$ is constructed by choosing suitable lapse and the shift. Notably, if the longitudinal potential of the shift is chosen such that equation \eqref{eq: shift_chi} holds, the two-spheres are immediately guaranteed to form an inverse mean curvature foliation in  $(\Sigma,h_{ij})$. We showed then that if on a ${\rho}=const$ level set the area and lapse satisfy the integral expression \eqref{eq: int-MG} then the non-negativity of the Geroch mass is guaranteed on that level set. If the constructed three-space happens to be asymptotically flat, mild integral conditions guarantee that the positive energy theorem and the Penrose inequality hold. We also pointed out that the quasi-local mass introduced by Bartnik in the context of quasi-spherical foliations could also be viewed as a special case of the Geroch mass.

\medskip 

It is important to emphasize that no assumption was made anywhere in our analysis concerning the sign of the scalar curvature of the constructed three-spaces. The reader may doubt if \eqref{eq: prop2} or \eqref{eq: GQLM2-cond} hold then. The following example demonstrates that these relations hold even if the scalar curvature becomes negative on parts of the foliating two-surfaces.
\begin{example*}\label{ex: example}
Consider first a $t_{Schw}=const$ timeslice of a Schwarzschild spacetime of mass $M$. Such a $t_{Schw}=const$ timeslice spans from the bifurcation surface to spacelike infinity, and it is also foliated by the $r=const\,(\geq 2M)$ metric spheres. The latter is, indeed,  a quasi-spherical foliation such that ${}^{{}^{(3)}}\hskip-1mm R_{Schw}=0$, and such that ${\widehat \gamma}_{AB}=r^2\,{\interior \gamma}_{AB}$, $\widehat  N{}^A_{Schw}=0$ and ${\widehat N}_{Schw}= 1/\sqrt{1-2M/r}$. The desired Riemannian three-metric is constructed by adding the term  $c\,M^2\,r^{-2}\,Y_1{}^0$ to ${\widehat N}_{Schw}$, where $c$ is a constant and $Y_1{}^0$ stands for the spherical harmonics, with $\ell=1$ and $m=0$. Accordingly, the constructed Riemannian three-metric on $\Sigma$\,\footnote{Here $\Sigma$ is a shortcut for the $t_{Schw}=const$ timeslice.} is determined via the relations
	\begin{equation}\label{ex: lapse}
		{\widehat N}= {\widehat N}_{Schw} + c\,M^2\,r^{-2}\,Y_1{}^0\,,
	\end{equation}
${\widehat \gamma}_{AB}=r^2\,{\interior \gamma}_{AB}$ and $\widehat  N{}^A=0$. It is straightforward to verify then that on any of the $r=const$ level sets the scalar curvature vanishes at the equatorial $\theta=\pi/2$, and, if $c>0$,  ${}^{{}^{(3)}}\hskip-1mm R$ is positive and negative on the upper and lower hemispheres, respectively.  In particular, if $c=10^{-1}$, the integral
	\begin{equation}
		\int_{\mathscr{S}_r}\Big[{}^{{}^{(3)}}\hskip-1mm R  +  \interior{{\widehat K}}{}_{kl} \interior{{\widehat K}}{}^{kl} + 2\,{\widehat N}^{-2}\,{\widehat \gamma}^{kl}\,({\widehat D}{}_k {\widehat N}) ({\widehat D}_l {\widehat N})\Big]\,{\widehat{\boldsymbol{\epsilon}}}
	\end{equation} 
is negative on the interval $2M < r < r_*\approx 3.33318\cdot 2M$, implying that the Geroch mass decreases there. Nevertheless, as the lapse, ${\widehat N}$ in \eqref{ex: lapse}, is everywhere greater than one, in virtue of \eqref{eq: GQLM22}, the Geroch mass must remain positive throughout $\Sigma$. It is also important to note that the yielded Riemannian three-space is asymptotically flat with $M_{ADM}=M$. Note also that the $r=2M$ level set is a minimal surface on $\Sigma$ and that the inequality \eqref{eq: prop-int} holds. This implies that the conditions in Theorem\,\ref{theo: asympt} are satisfied, i.e., both the positive mass theorem and the Penrose inequality hold.
\end{example*}
An abundance of analogous examples may be created such that they all admit an inverse mean curvature foliation, such that, in the meanwhile, ${}^{{}^{(3)}}\hskip-1mm R \geq 0$ does not hold. Notably, these spaces cannot be part of maximal slices, and thereby, as already indicated, they are out of the validity range of the results covered in \cite{HusIlm97,HusIlm01,Bray01,Bray09}.

\medskip

It is also important to emphasize that a wide variety of Riemannian three-spaces can be constructed by the proposed method. Each of these spaces possesses a smooth inverse mean curvature foliation. To see this, recall first that in fixing the shift, only one of the potentials in \eqref{eq: hodge} gets to be determined by solving \eqref{eq: shift_chi}, whereas the other potential remains freely specifiable. Note also that, as far as only the construction of the inverse mean curvature foliations is concerned, in virtue of Corollary \ref{corolary1}, the topology of $\Sigma$ can be allowed to be generic. Note, finally, that as the choice made for the lapse does not affect the inverse mean curvature character of the foliation, the corresponding freedom also enlarges the variety of the constructed three-spaces.  

\medskip

To give another, possibly more tempting, application of the proposed new construction, for simplicity, consider a three-manifold $\Sigma$ with topology $\mathbb{R}\times \mathbb{S}^2$, i.e.~it is  smoothly foliated by topological two-spheres (determined by a smooth function $\rho: \Sigma \rightarrow \mathbb{R}$). Assume that $h_{ij}$ is a smooth three-metric on $\Sigma$. Choose, as described in section 2, a smooth flow on $\Sigma$. If adapted coordinates are used, the metric takes the form (3.19) in the pertinent coordinate patch determined by the triplet $({\widehat N}\,, {\widehat N}{}^A\,, {\widehat \gamma}_{AB})$. The $\rho=const$ level sets in $\Sigma$, in general, have no chance to form an inverse mean curvature foliation of the Riemannian three-space $(\Sigma,{h}_{ij})$. This is so because, as noted at the end of subsection 3.1, whenever the three-metric ${h}_{ij}$ on $\Sigma$ is fixed, the entire geometric setup gets to be too rigid. Nevertheless, the construction proposed in this paper is guaranteeing that the very same foliation gets to be an inverse mean curvature foliation for the Riemannian three-metric $h^*_{ij}$ on $\Sigma$ that is yielded from $h_{ij}$ by replacing the lapse ${\widehat N}$ and shift ${\widehat N}{}^A$ by suitably chosen new lapse ${\widehat N}_*$ and shift ${\widehat N}{}_*^A$ while keeping the two-metric ${\widehat \gamma}_{AB}$, the $\rho=const$ level sets and the flow vector $\rho^i$ intact. Notice that the fixed elements in $\{  \,\rho: \Sigma \rightarrow \mathbb{R}\,, \rho^i=(\partial_{\rho})^i\,;\,{\widehat \gamma}_{AB} \}$ can be used to evaluate both of the Lie derivatives on the right hand side of \eqref{eq: shift1} without referring to any additional geometric structure. The construction comes into play by solving \eqref{eq: shift_chi} for the potential $\chi$. It is important to emphasize that neither the shift ${\widehat N}{}_*^A$ nor the three-metric $h^*_{ij}$ is known yet. Nevertheless, for a great surprise, we do already have the guarantee that, regardless of choice we shall make for the other, freely specifiable, potential $\eta$ in \eqref{eq: hodge} and the new lapse ${\widehat N}_*$, the relation ${{\widehat{N}}_* {{\widehat K}_*}{}^l{}_{l}}=\overline{{\widehat{N}_*} {{\widehat K}_*}{}^l{}_{l}}$ will hold. This may be really astonishing as neither ${\widehat{N}}_*$ nor ${{\widehat K}_*}{}^l{}_{l}$, appearing in this relation, are known yet. Nevertheless, things are in order as the unit normal $\widehat{n}_*{}^i$ and, in turn, ${{\widehat K}_*}{}^l{}_{l}$, gets to be determined via the relations  
\begin{equation}
\widehat{n}_*{}^i= {\widehat{N}_*}^{-1}\left[(\partial_{\rho})^i-{\widehat N}{}_*^A\,(\partial_{A})^i \right]\,, \quad {\rm and} \quad {{\widehat K}_*}{}^l{}_{l}= D_i {\widehat n}_*{}^i 
\end{equation}
only after, in addition to the other potential $\eta$, fixing the shift ${\widehat N}{}_*^A$ via \eqref{eq: hodge}, and the new lapse ${\widehat N}_*$ gets also to be fixed. 

It is an interesting issue how close the two metrics, $h_{ij}$ and $h^*_{ij}$, can be placed to each other in the space of three-metrics on $\Sigma$. Given a  Riemannian three-space $(\Sigma,h_{ij})$ is it possible to construct a three-metric, analogous to $h^*_{ij}$, on $\Sigma$ such that a suitably chosen smooth foliation of $(\Sigma,h_{ij})$ gets to be an inverse mean curvature foliation for the constructed Riemannian three-space $(\Sigma,h^*_{ij})$ and such that in the mean time $h^*_{ij}$ is close to $h_{ij}$? This question is leading out of the scope of the present paper, and it is left for future investigations.

\medskip

It is also an interesting question what would happen if the requisite conditions employed to show the non-negativity of the Geroch mass and Penrose inequality were not satisfied. For example, what kinds of spaces violate (5.31)? Are the corresponding spaces pathological? Though the answer to these is not known, the author would like to recall that the principal aim of the present paper is to introduce a construction of Riemannian three-geometry with IMCF. It is also worth emphasizing that the lapse $\widehat N$ in the proposed construction is freely specifiable, and its choice does not affect the IMC character of the constructed foliation. Thereby, one could choose the lapse such that, e.g. (5.31) is violated. Whether the corresponding Riemannian three-geometry is pathological or not depends on the context fixed by some additional yet unspecified conditions. Thereby, the questions raised above are out of the present paper's scope though they would deserve further investigations.

\medskip

Note also that our main results are local in the radial direction. This would allow addressing various interesting questions concerning global properties. For instance, one may ask what conditions would guarantee asymptotic flatness of the constructed Riemannian three-spaces. Though interesting, this problem is out of the scope and is left for later investigations.

\medskip

Note, finally, that merely the Riemannian character of the two-metric $\widehat{\gamma}_{AB}$ on the $\mathscr{S}_\rho$ level sets, along with that of the three-metric $h_{ij}$ on $\Sigma$, was assumed. Accordingly, if $\Sigma$ happens to be a hypersurface in some four-dimensional ambient space, the signature of the pertinent four-metric,  $g_{ab}$, can be either Lorentzian or Euclidean. Note also that no field equation restricting the metric $h_{ij}$ on $\Sigma$ or the four-metric $g_{ab}$ on some four-dimensional ambient space was used anywhere in our analysis. An immediate implication of these observations is that the proposed construction applies essentially to Riemannian three-spaces in any metric theory of gravity.

\section*{Acknowledgments}
 
This project was supported in part by the POLONEZ programme of the National Science Centre of Poland (under the project No. 2016/23/P/ST1/04195) which has received funding from the European Union`s Horizon 2020 research and innovation programme under the Marie Sk{\l}odowska-Curie grant agreement No.~665778 and by the NKFIH grant K-115434. 
\includegraphics[scale=0.22, height = 12pt]{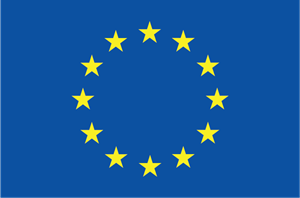}

\section*{Data Availability}

Data sharing not applicable to this article as no datasets were generated or analyzed during the current study.


\end{document}